\documentclass[useAMS,usenatbib]{mn2e}
\usepackage{epsfig}
\usepackage{threeparttable}
\usepackage{amsmath}

\title[An evolutionary channel towards SAX J1808.4-3658]{An evolutionary channel towards the accreting millisecond pulsar SAX J1808.4-3658}
\author[W. C. Chen]{ Wen-Cong Chen $^{1,2}$\thanks{E-mail:
chenwc@pku.edu.cn}\\
$^1$ School of Physics and Electrical Information, Shangqiu Normal University,
Shangqiu 476000, China;\\
$^2$ Department of Astrophysics, University of Oxford, Oxford OX1 3RH, UK;\\
 }
\begin{document}

\date{}

\pagerange{\pageref{firstpage}--\pageref{lastpage}} \pubyear{2011}

\maketitle

\label{firstpage}

\begin{abstract}
Recent timing analysis reveals that the orbital period of the first discovered accreting millisecond pulsar SAX J1808.4-3658 is increasing at a rate $\dot{P}_{\rm orb}=(3.89\pm0.15)\times 10^{-12}~\rm s\,s^{-1}$, which is at least
one order of magnitude higher than the value arising from the conservative mass transfer.
An ejection of mass loss rate of $10^{-9}~\rm M_{\odot}{\rm yr}^{-1}$ from the donor star at the inner Lagrangian point during the quiescence state could interpret the observed orbital period derivative. However, it is unknown whether this source can offer such a high mass loss rate. In this work, we attempt to investigate an evolutionary channel towards SAX J1808.4-3658. Once the accretion disk becomes thermally and viscously unstable, the spin-down luminosity of the millisecond pulsar and the X-ray luminosity during outbursts are assumed to evaporate the donor star, and the resulting winds carry away the specific orbital angular momentum at the inner Lagrangian point. Our scenario could yield the observed orbital period, the orbital period derivative, the peak X-ray luminosity during outbursts. Low-mass X-ray binaries with a $1.0~\rm M_{\odot}$ donor star, and an orbital period in the range of 0.8 - 1.5 d, may be the progenitor of SAX J1808.4-3658. Our numerical calculations propose that the current donor star mass is $0.044 ~\rm M_{\odot}$, which is in approximately agreement with the minimum mass of the donor star. In addition, our scenario can also account for the formation of black widows or the diamond planet like PSR J1719-1438.
\end{abstract}


\section{Introduction}
It is generally accepted that binary millisecond pulsars evolved from low-mass X-ray binaries (LMXBs) via the recycling channel, in which the pulsar accreted the material and angular momentum from the donor star, and is spun up to a millisecond period \citep{alpa82,bhat91}. In this scenario, accreting millisecond pulsars (AMSPs) are very important evolutionary link between LMXBs and binary millisecond pulsars. Since the first AMSP SAX J1808.4-3658 has been discovered \citep{wijn98}, the number of this population has already exceeded fourteen. As the first AMSP, SAX J1808.4-3658 is a vital fossil uncovering the evolutionary history of LMXBs.

Same to other AMSPs, SAX J1808.4-3658 is also a transient X-ray source. As the first discovered AMSP, SAX J1808.4-3658 has been extensively studied in observations. In 1996 September, two very bright type I X-ray bursts from this source were observed by the BeppoSAX Wide Field Cameras \citep{int98,int01}. The subsequent observation by \emph{ RXTE/PCA} reveals $\sim 2.5$ ms X-ray pulsation period and $\sim 2$ hours orbital period \citep{chak98,wijn98}. From 1998 to 2015, observations by \emph{RXTE/PCA} and \emph{XMM Newton} with high time resolution have detected several outbursts \citep[recent observations see also][]{patr12,sann15}, which have recurrence intervals of 1.6 to 3.3 years. Its bolometric luminosity is $L_{\rm bol}\la 10^{32}~\rm erg\,s^{-1}$ during quiescence \citep{camp02,hein07}, and the peak bolometric luminosity is $L_{\rm bol}= (5-8)\times10^{36}~\rm erg\,s^{-1}$ \citep{hart08}.

A timing analysis of four X-ray bursts from SAX J1808.4-3658 during 1998$-$2005 indicated that its spin frequency $\nu$ is decreasing at a rate $\dot{\nu} =(-5.6\pm2.0)\times10^{-16}\rm s^{-2}$, and the orbital period is increasing at a rate  $\dot{P}_{\rm orb}=(3.5\pm0.2)\times 10^{-12}~\rm s\,s^{-1}$ during quiescence \citep{hart08,di08}. Based on XMM-Newton observation for the fifth outburst, \cite{burd09} derived an orbital period derivative $\dot{P}_{\rm orb}=(3.89\pm0.15)\times 10^{-12}~\rm s\,s^{-1}$. This parameter is at least one order of magnitude higher than the value arising from the conservative mass transfer. Based on the consequential angular momentum loss mechanism \citep{king95}, the observed orbital period derivative may originated from a $\sim10^{-9}~\rm M_{\odot}{\rm yr}^{-1}$ mass loss rate, which was ejected by the radiation pressure of the pulsar from the inner Lagrangian point \citep{di08,burd09}.  As an alternative model, \cite{hart09} proposed that the orbital period change of this source is likely a short-term event, in which the magnetic activity of the donor star would result in an exchange of angular momentum between the donor star and the orbit caused by the tidal lock \citep{appl92,appl94}. Applegate mechanism, however, in most cases produces an orbital period change in less than 10 yr, in contradiction with what is observed in SAX J1808.4-3658.

In a word, the relatively large orbital period derivative of SAX J1808.4-3658 still remains mysterious so far. To explain the existence of AMSPs such as SAX J1808.4-3658, we have found an evolutionary channel to understand its formation process. Based on the accretion disk instability model, the evaporation wind and the irradiation wind model, in this work we attempt to investigate the formation process of SAX J1808.4-3658. In section 2, we analyze the orbital period evolution of SAX J1808.4-3658. In section 3, we performed a numerical calculation for SAX J1808.4-3658. Finally, we summarize the results with a brief discussion in section 4.

\section{Analysis on the orbital period evolution of SAX J1808.4-3658}
The orbital angular momentum of a binary system is $J=M_{\rm NS}M_{\rm d}/(M_{\rm NS}+M_{\rm d})a^{2}\Omega$, where $a$  is the orbital separation, $\Omega$ the orbital angular velocity of the binary, $M_{\rm NS}$, and $M_{\rm d}$ are the pulsar mass, and the donor star mass, respectively. Differentiating this equation, we have the change rate of orbital period as follows
\begin{equation}
\frac{\dot{P}_{\rm orb}}{P_{\rm orb}}=3\frac{\dot{J}}{J}-3\frac{\dot{M}_{\rm d}}{M_{\rm d}}(1-q\beta)+\frac{\dot{M}_{\rm NS}+\dot{M}_{\rm d}}{M_{\rm NS}+M_{\rm d}},
\end{equation}
where $\beta=-\dot{M}_{\rm NS}/\dot{M}_{\rm d}$ is the ratio between the accretion rate of the pulsar and the mass loss rate of the donor star, $q=M_{\rm d}/M_{\rm NS}$ is the mass ratio of the binary. In this section, we consider the angular momentum loss by the gravitational radiation and the donor star winds, i. e. $\dot{J}=\dot{J}_{\rm gr}+\dot{J}_{\rm w}$.

Taking $M_{\rm NS}=1.4~\rm M_{\odot}$, and $P_{\rm orb}=0.0839~\rm d$, the angular momentum loss rate of SAX J1808.4-3658 by gravitational radiation can be written as
\begin{equation}
\frac{\dot{J}_{\rm gr}}{J}=-8.68\times 10^{-17}\frac{m_{\rm d}}{(1.4+m_{\rm d})^{1/3}}~\rm s^{-1},
\end{equation}
where $m_{\rm d}$ is $M_{\rm d}$ in units of solar mass. The mass loss rate of the donor star is $\dot{M}_{\rm d}=\dot{M}_{\rm tr}+\dot{M}_{\rm wind}$, where $\dot{M}_{\rm tr}$, $\dot{M}_{\rm wind}$ are the mass transfer rate, and the wind loss rate, respectively. Assuming $\dot{M}_{\rm tr}=-\dot{M}_{\rm NS}$, the angular momentum loss rate by the donor star wind satisfies the following formula
\begin{equation}
\frac{\dot{J}_{\rm w}}{J}=\frac{\alpha(1-\beta)\dot{M}_{\rm d}}{(1+q)M_{\rm d}},
\end{equation}
where $\alpha$ is the specific angular momentum of the wind loss in units of that of the donor star. If the donor star wind was ejected at the inner Lagrangian point, $\alpha\approx(1-0.462(1+q)^{2/3}q^{1/3})^{2}\approx0.7$ \citep{di08} based on the approximate Roche lobe radius given by \cite{pacz71}.

Inserting equation (3) into equation (1), one can yield an equation that is similar to equation (3) of \cite{di08} as follows
\begin{equation}
\frac{\dot{P}_{\rm orb}}{P_{\rm orb}}=3\frac{\dot{J}_{\rm gr}}{J}-3\frac{\dot{M}_{\rm d}}{M_{\rm d}}\left(1-q\beta-(1-\beta)\frac{\alpha+q/3}{1+q}\right).
\end{equation}

To account for the large orbital period derivative, SAX J1808.4-3658 was proposed to be experience a non-conservative mass transfer.
Setting $\beta=0$, the wind loss rate of the donor star is given by
\begin{equation}
\dot{M}_{\rm d}=\frac{M_{\rm d}(\frac{\dot{J}_{\rm gr}}{J}-\frac{\dot{P}_{\rm orb}}{3P_{\rm orb}})}{1-\frac{\alpha+q/3}{1+q}}.
\end{equation}
Inserting the observed parameter $\dot{P}_{\rm orb}/P_{\rm orb}=5.4\times 10^{-16}~\rm s^{-1}$, we can plot the relation between the mass loss rate and the mass of the donor star in Figure 1. For a conservative mass transfer, the mass transfer rate $(2-3)\times10^{-10} \rm M_{\odot}{\rm yr}^{-1}$ in the range of 0.04-0.05 $\rm M_{\odot}$ donor star mass would yield an X-ray luminosity of $\sim(1.4-2)\times10^{36}~\rm erg\,s^{-1}$, which is 1-2 orders of magnitude higher than the averaged X-ray luminosity of $\sim 4\times10^{34}~\rm erg\,s^{-1}$ \citep{di08}. Conversely, there exist no X-ray luminosity contradiction for the nonconservative mass transfer. When $\alpha=1$ ,  $\dot{M}_{\rm d}\approx\frac{3(1+q)M_{\rm d}}{2q}\frac{\dot{P}_{\rm orb}}{3P_{\rm orb}}\approx(1+q)M_{\rm NS}\frac{\dot{P}_{\rm orb}}{2P_{\rm orb}}\approx1.2\times 10^{-8}\rm M_{\odot}{\rm yr}^{-1} $, the mass loss rate of a low-mass donor star is independent of its mass (see also the solid curve in Figure 1). To produce the observed orbital period derivative, a mass loss rate of $\sim 10^{-9} \rm ~M_{\odot}{\rm yr}^{-1}$ with the specific angular momentum at the inner Lagrangian point could be a choice for SAX J1808.4-3658 \citep{di08}.

\begin{figure}
\centering
\includegraphics[width=\linewidth,trim={30 0 90 30},clip]{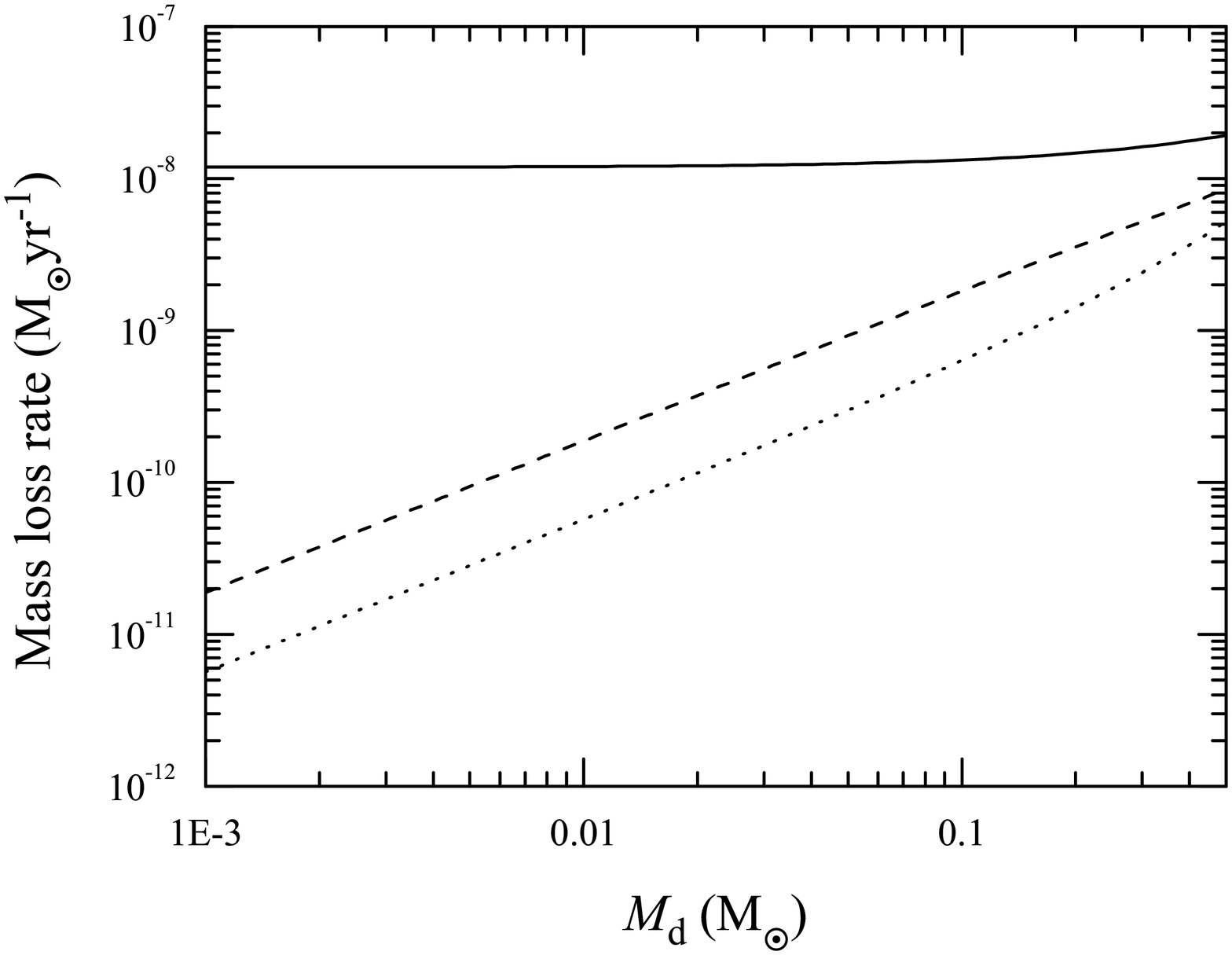}
\caption{Mass loss rate vs. the donor star mass of SAX J1808.4-3658. The solid, and dashed curves represent the non-conservative mass transfer scenario ($\beta=0$) that the lost matter carries away the specific angular momentum of the donor star ($\alpha=1$), and the inner Lagrangian point ($\alpha=0.7$), respectively. The dotted curve indicates the conservative mass transfer case ($\beta=1$).} \label{fig:orbmass}
\end{figure}

\section{Evolutionary example toward SAX J1808.4-3658}

\subsection{Evolution code}
In this section, employing \texttt{MESAbinary} version (7624)
in \texttt{MESA} module \citep{paxt11,paxt13,paxt15}, we attempt to simulate the formation process of SAX J1808.4-3658.  The evolutionary beginning is a binary system containing a low-mass donor star (with a mass of $M_{\rm d} = 1.0 ~\rm M_{\odot}$) and a neutron star (with a mass of $M_{\rm NS} = 1.4 ~\rm M_{\odot}$), and with a circular orbit. For the donor star compositions, we adopt a solar compositions ($X = 0.70, Y = 0.28$, and $Z = 0.02$).

During the mass transfer, we take an accretion efficiency of 0.5 (this value is a bit arbitrary) similar to \cite{pods02}. Meanwhile, the accretion rate of the neutron star is limited by the Eddington accretion rate $\dot{M}_{\rm Edd}$, i. e. $\dot{M}_{\rm NS}={\rm min}(-\dot{M_{\rm tr}}/2,\dot{M}_{\rm Edd})$.  In addition, if the mass transfer rate is lower than the critical mass transfer rate \citep{para96,dubu99}
\begin{eqnarray}
\dot{M}_{\rm cr} \simeq
3.2\times10^{-9}\left(\frac{M_{\rm NS}}{1.4M_{\odot}}\right)^{0.5}
\left(\frac{M_{\rm d}}{1M_{\odot}}\right)^{-0.2} \nonumber\\
\left(\frac{P_{\rm orb}}{1 \rm d}\right)^{1.4}\quad \rm M_{\odot}{\rm yr}^{-1},
\end{eqnarray}
the accretion disk should experience a thermally and viscously instability. During this stage, the accreting neutron star would be observed as a transient X-ray source, which appears short-lived outbursts separated by long-term quiescence. The neutron star was assumed to accrete at a rate of $\dot{M}_{\rm NS}={\rm min}(-\dot{M}_{\rm tr}/(2d),\dot{M}_{\rm Edd}$) (where $d$ is the duty cycle, and it defined by the ratio between the outburst timescale
and the recurrence time) during outbursts, and almost no accretion occurs during quiescence. The mass loss during accretion (including the super-Eddington accretion in both persistent state and outbursts duration in transient state) is assumed to form an isotropic wind in the vicinity of the neutron star.

To account for the period derivative of SAX J1808.4-3658, we consider the influence of the evaporation wind driven by the spin-down luminosity of the millisecond pulsar. In calculation, evaporation would work once the LMXB satisfied the following two criterions \citep{jia15}: (1) the neutron star has already evolved into a millisecond pulsar after it accreted $0.1\rm ~ M_{\odot}$; (2) the accretion disk instability occurs, the neutron star appears as a transient, and is spinning down by magnetic dipole radiation during the quiescence state \footnote{Once the accretion disk instability takes place, the neutron star accretes at an ultra-low accretion rate during the quiescent phase. Because the magnetosphere radius $r_{\rm m}\propto \dot{M}_{\rm NS}^{-2/7}$, it would be outside the light cylinder. Therefore, the pulsar was thought to be in 'radio ejection' phase, and prevent further accretion \citep{burd01,burd02}.  The magnetic dipole radiation during outbursts may be reduced or shielded by the abundant material around the pulsar.}. The wind loss rate of the donor star by evaporation process is given by \citep{heuv88,stev92}
\begin{equation}
\dot{M}_{\rm ev}=-\frac{f_{\rm ev}}{2v^{2}_{\rm esc}}L_{\rm p}\left(\frac{R_{\rm d}}{a}\right)^{2},
\end{equation}
where $f_{\rm ev}$ is the evaporation efficiency, $v_{\rm esc}$ is the escape velocity at the donor star surface, $R_{\rm d}$ is the donor star radius. Recently, \cite{bret13} obtained a 10 - 30 \% efficiency that the spin-down luminosity of pulsars were used to heat on the surface of the donor star. To account for the large orbital period derivative of SAX J1808.4-3658, we take $f_{\rm ev}=0.3$. The spin-down luminosity of the pulsar is given by
$L_{\rm p}=4\pi^{2}I\nu\dot{\nu}$ (where $I$ is the momentum of inertia of the pulsar), and the spin-down evolution was assumed to obey a power law $\dot{\nu}=-K\nu^{3}$ (where $K$ was assumed to be a constant depending on the momentum of inertia, the magnetic field, and the radius of the pulsar). In numerical calculation, we take $\nu=401~\rm s^{-1}$, $\dot{\nu}=-5.5\times 10^{-16}~\rm s^{-2}$ \citep{hart08}, and $I=10^{45}~\rm g\,cm^{2}$, so $K=-\dot{\nu}/\nu^{3}=8.5\times 10^{-24}~\rm s$. Furthermore, a duty cycle $d=0.01$ is adopted. Due to a relatively short duration and a high spin frequency, we neglect the spin-up of the pulsar during outbursts.

In addition, once LMXBs appears to be a transient, we also consider the irradiation-driving wind by X-ray luminosity during outbursts. The mean irradiation-driving wind loss rate is given by
\begin{equation}
\dot{M}_{\rm ir}=-d \times f_{\rm ir}L_{\rm X}\frac{R_{\rm d}^{3}}{4 G M_{\rm d} a^{2}},
\end{equation}
where $f_{\rm ir}$ is the irradiation efficiency, and the wind loss rate of the donor star $\dot{M}_{\rm wind}=\dot{M}_{\rm ev}+\dot{M}_{\rm ir}$. We calculate the outburst X-ray luminosity by $L_{\rm X}=0.1\dot{M}_{\rm NS}c^{2}$, and take a relatively high irradiation efficiency $f_{\rm ir}=0.1$.

\begin{figure}
\centering
\includegraphics[width=\linewidth,trim={30 0 90 30},clip]{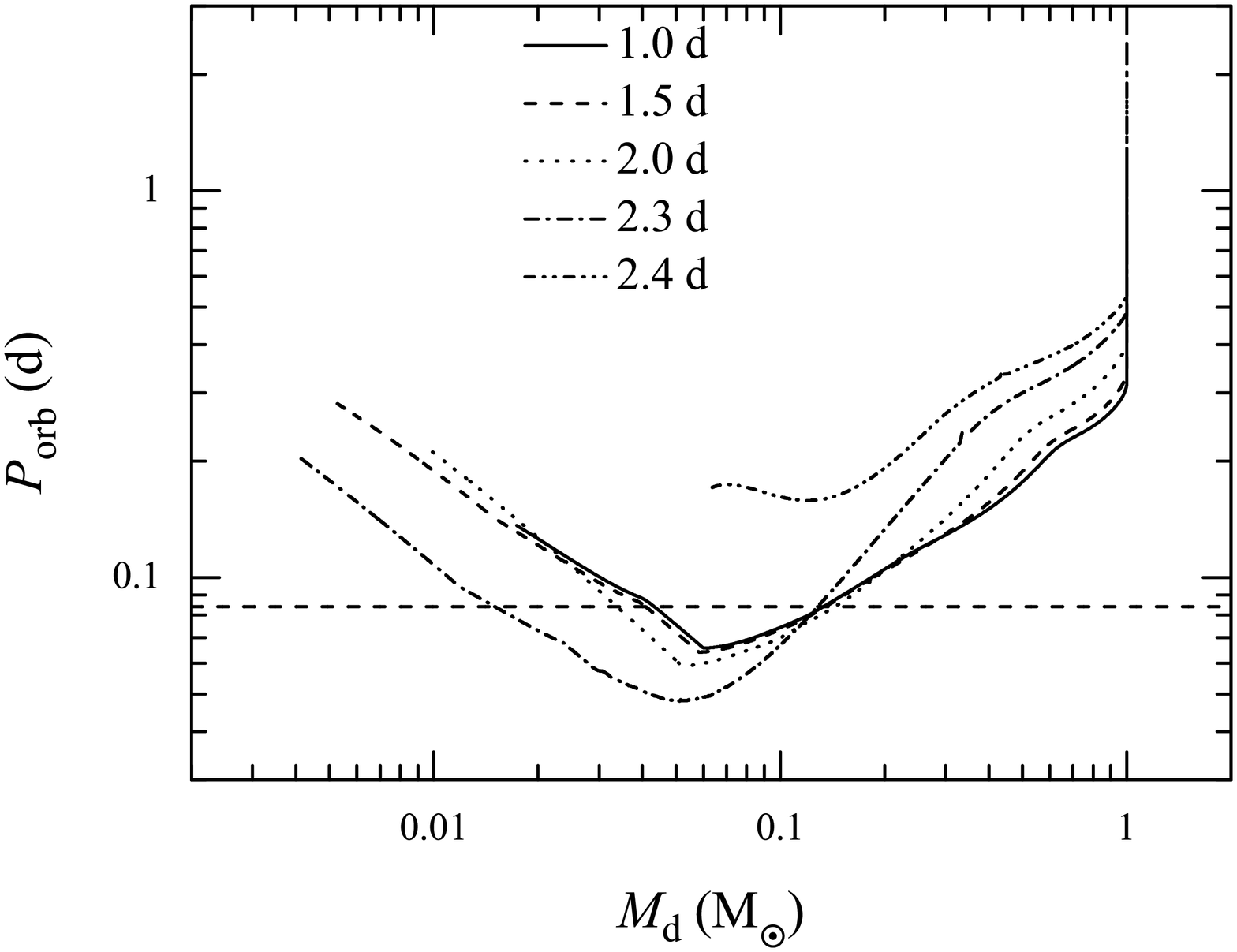}
\caption{Evolutionary tracks of LMXBs with a donor star mass of $1.0~\rm M_{\odot}$ and different initial orbital periods in the $P_{\rm orb} - M_{ \rm d}$ diagram. The horizontal dashed line denotes the observed orbital period of SAX J1808.4-3658.} \label{fig:orbmass}
\end{figure}

In the \texttt{MESA} code, we consider three types of orbital angular momentum
loss during the evolution of LMXBs: (1) gravitational-wave radiation; (2)
magnetic braking: we adopt the standard magnetic braking prescription
given by \cite{rapp83}, and take $\gamma=4$; (3) mass loss: the
mass loss from the vicinity of the pulsar is
assumed to carry away the specific orbital angular
momentum of the pulsar \citep{taus99}, while the donor star winds was assumed to be ejected from the inner Lagrangian point by the radiation pressure of the pulsar \citep{burd01}, and carry away the specific orbital angular momentum at this point \citep{burd02,jia15}.

Table 1 summarizes the calculated results when the magnetic braking cut-off is included.  In the case that initial orbital-periods is $P_{\rm orb,i}=2.2$ d, a low mass transfer rate (less than $\dot{M}_{\rm cr}$) due to the sudden stop of magnetic braking would result in an accretion disk instability. The subsequent evaporation and irradiation process produce a high wind loss rate, and give rise to a detached binary system (according to the second term on the right side of equation 1, mass loss from the donor star would cause an orbital expansion), so the binary could not become a transient X-ray source. Due to $P_{\rm orb,i}=2.3$ d is near the bifurcation period (see also section 3.2), the binary can reach a lowest minimum orbital-period. Strong gravitational radiation can induce the donor star in a detached binary to fill its Roche lobe again. However, the donor star mass and the mass transfer rate are not consistent with the observed data. Therefore, we keep the magnetic braking on even if the donor star loses its radiative core. Magnetic braking can play a key role to keep a semi-detached binary under such a strong wind loss.
\begin{table}
\begin{center}
\caption{Calculated results of LMXBs with a $1.0~\rm M_{\odot}$ donor star and various initial orbital-period. The meaning of the columns are presented as follows: the initial orbital-period; the minimum orbital-period; and the orbital-period derivative, the mass transfer rate, and the donor star mass when the orbital-period can fit that of SAX J1808.4-3658. In calculation, the magnetic braking is not included when the donor stars lose their radiative core. \label{tbl-1}}
\begin{tabular}{ccccc}
\hline
$P_{\rm orb,i}$ & $P_{\rm orb,min}$ &  $ \dot{P}_{\rm orb}$&  $\dot{M}_{\rm tr}$ & $M_{\rm d}$\\
(d)&(d)&($\rm s\,s^{-1}$)&($\rm M_{\odot}\,yr^{-1}$) &($\rm M_{\odot}$) \\
\hline
2.4 & 0.158  & no   & no  & no   \\
2.3 & 0.050  & $5.9\times10^{-11}$  & $3.2\times10^{-10}$  &0.017 \\
2.2 & 0.058  & $4.6\times10^{-13}$  & 0  &0.034 \\
2.0 & 0.092 & no & no  & no  \\
1.5 & 0.114 & no & no  & no  \\
\hline
\end{tabular}
\end{center}
\end{table}

\subsection{Results}

 To investigate the orbital evolution process of SAX J1808.4-3658, we plot the evolutionary tracks of LMXBs with different initial orbital periods in the $P_{\rm orb} - M_{\rm d}$ diagram (see also Figure 2). To reach an ultra-short orbital period, the initial orbital period of LMXBs should be less than the so-called bifurcation period \citep{pods02}. Bifurcation period defined as the longest initial orbital period that could form ultra-compact X-ray binaries within a Hubble time \citep{sluy05a,sluy05b}, and it strongly depends magnetic braking efficiency and the mass loss of binary systems. For LMXB with a $1.0~\rm M_{\odot}$ mass donor star, our evolutionary scenario yield a bifurcation of $\approx2.3$ d. As shown in Figure 2, an initial orbital period near the bifurcation period could readily attain the shorter minimum orbital period.

\begin{figure}
\centering
\includegraphics[width=\linewidth,trim={30 0 40 30},clip]{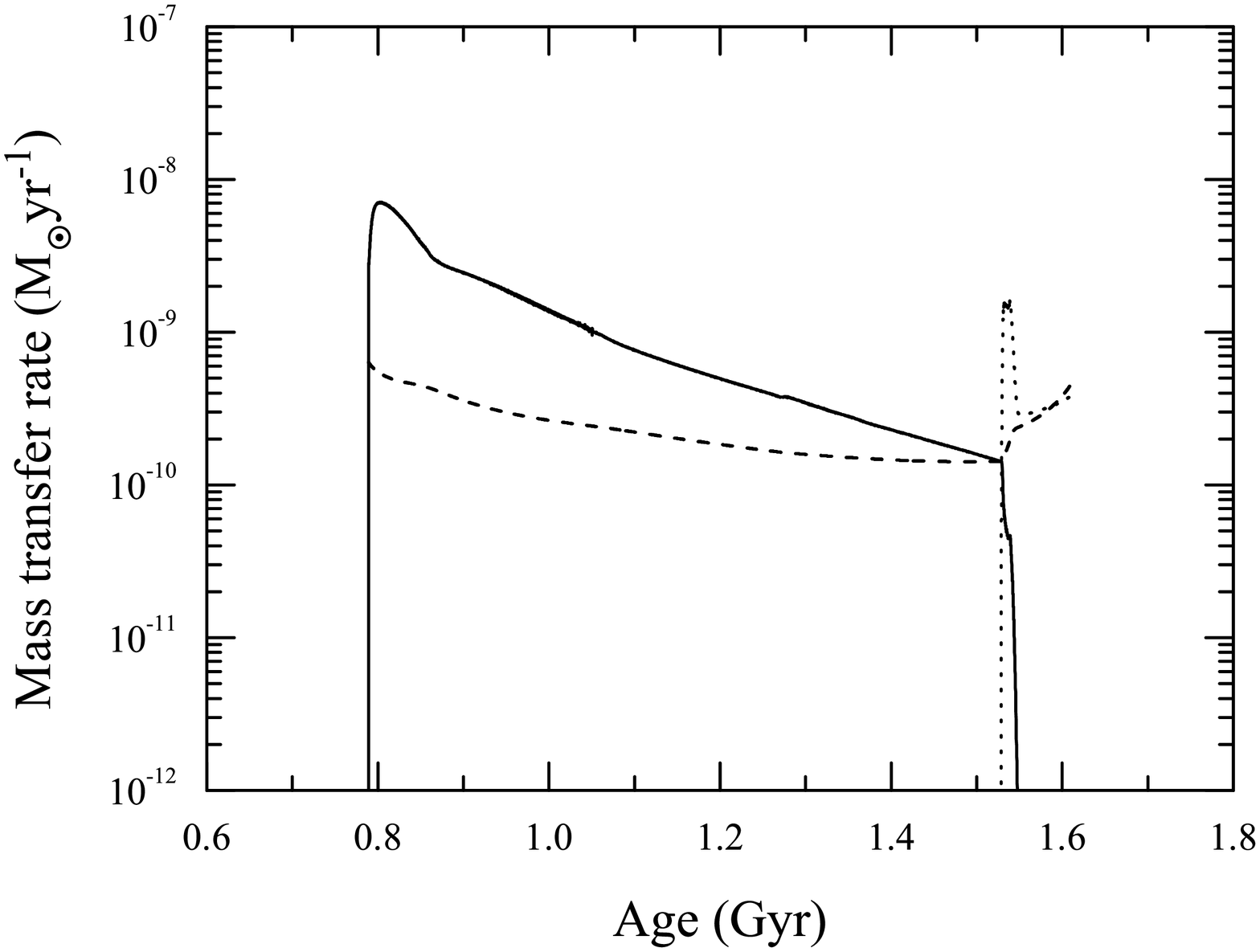}
\includegraphics[width=\linewidth,trim={30 0 40 30},clip]{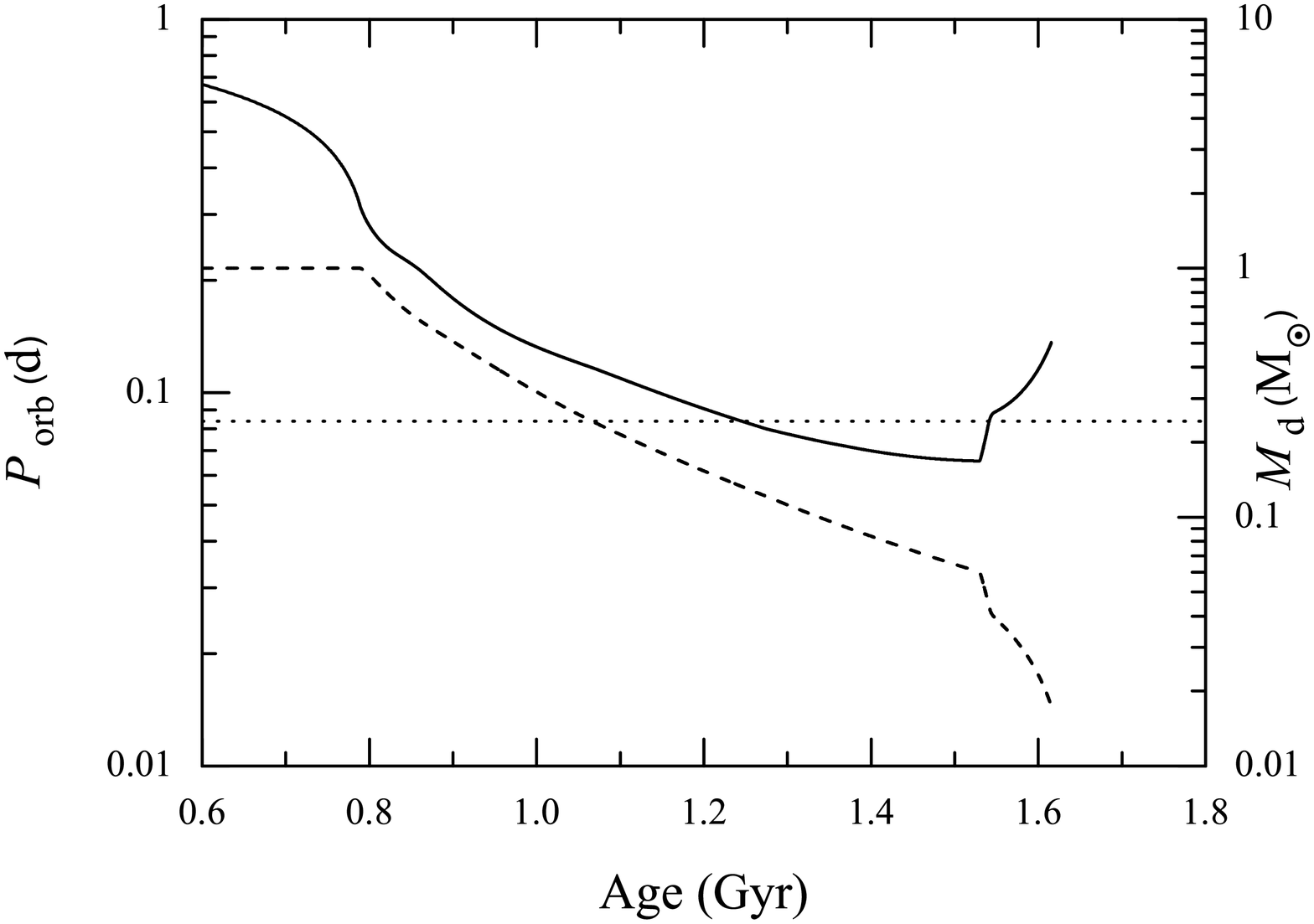}
\caption{Evolution of main binary parameters for an LMXB with a donor star mass of $1.0~\rm M_{\odot}$ and an initial orbital period of 1.0 days. Top panel: the mass transfer rate (solid curve), the critical mass transfer rate (dashed curve), and the wind loss rate of the donor star (dotted curve). Bottom panel: the orbital period (solid curve), the donor star mass (dashed curve), and   the orbital period (0.0839 d) of SAX J1808.4-3658 (dotted curve).} \label{fig:orbmass}
\end{figure}

Figure 3 plots the evolutionary sequences of an LMXB with a donor star mass of $1~\rm M_{\odot}$ and an initial orbital period of 1.0 d. Since the initial orbit is relatively compact, the donor star only underwent a short timescale nuclear evolution.
When $t=0.79$ Gyr, the donor star overflows its Roche lobe, and begins a Case A mass transfer. Due to the strong magnetic braking, the orbital period has already decreased to 0.31 d at the moment of Roche lobe overflow. The central H abundance remains a high fraction, $X_{\rm c}=0.64$, and the donor star is still in main sequence stage. The mass transfer firstly occurs at a rate of $(2.0-7.0)\times10^{-9}~ \rm M_{\odot}yr^{-1}$. At the age of 1.53 Gyr, the mass transfer rate declines to below the critical mass transfer rate, and evaporation and irradiation processes begin. At this moment, LMXB evolves to a minimum orbital period of 1.57 hr. Because of the accreted mass $\bigtriangleup M_{\rm acc} = 0.47~ \rm M_{\odot}$, the neutron star can be spun up to $\approx 2 - 3$ ms (see also Figure 5 of Liu \& Chen 2011). \cite{shib89} presented an empirical formula of magnetic field evolution during the accretion
\begin{equation}
B=\frac{B_{\rm i}}{1+\bigtriangleup M_{\rm acc}/m_{\rm B}},
\end{equation}
 where $B_{\rm i}$ is the initial magnetic field of the neutron star, and $m_{\rm B}\sim 10^{-4}\rm M_{\odot}$ is a parameter describing the magnetic field decay. Assuming $B_{\rm i}=10^{12}$ G, $B\approx 2\times 10^{8}$ G, in approximately agreement with the estimation ($\sim3.5\pm 0.5\times 10^{8}$ G) given by \cite{burd06}. Subsequently, the mass transfer rate begins to sharply decrease to $\sim10^{-12}-10^{-11}~ \rm M_{\odot}yr^{-1}$  because of the orbital expansion caused by the strong evaporation and irradiation process. At 1.56 Gyr, the binary becomes a detached system, and will not appear as a strong X-ray source. Both in the period-decreasing and period-increasing stages, the binary system can attain the observed orbital period (0.0839 d) of SAX J1808.4-3658. However, SAX J1808.4-3658 should be in the period-increasing phase at present according to the observation. At $P_{\rm orb}=0.0839~\rm d$, the angular momentum loss rates by gravitational radiation, magnetic braking, and mass loss are $-1.5\times 10^{33}$, $-1.2\times 10^{34}$, and $-2.2\times 10^{35}~\rm g\,cm^{2}\,s^{-2}$, respectively. The angular momentum loss by mass loss is stronger than that by magnetic braking. However, the mass transfer would not continue if we turn off the magnetic braking.

In Figure 4, we illustrate the mass loss rates induced by angular momentum losses caused by gravitational radiation (mass loss rate has been calculated by equation 5 of King et al. 2005), irradiation process, and evaporation process. With the beginning of evaporation and irradiation process, the orbit of the binary broadens, and the mass transfer rate induced by gravitational radiation sharply declines. Even if we consider magnetic braking for a fully convective star, the mass transfer rate is obviously less than $\sim10^{-9}~ \rm M_{\odot}yr^{-1}$ when the donor star mass is less than $0.1~ \rm M_{\odot}$. Since the equation for the mass transfer rate is not linear in the angular momentum
losses (see e.g. equation 6 of Di Salvo et al. 2008), therefore the total mass transfer rate is not simply the sum of the contributions to mass transfer rate due to gravitational radiation, magnetic braking, and the donor star wind loss.

\begin{table*}
\centering
\begin{minipage}{170mm}
\caption{Calculated relevant parameters of LMXBs when their orbital period can fit the observed values of SAX J1808.4-3658 in the period-increasing phase for different initial orbital periods. The meaning of the columns are presented as follows: initial orbital period, age, the donor star and the pulsar masses, orbital period, orbital period derivative, mass transfer
rate, wind loss rate, donor star radius, effective temperature, surface H abundance, and the mean density of the donor star.}
\begin{tabular}{llllllllllll}
  \hline\noalign{\smallskip}
 $P_{\rm orb,i}$ & Age   & $M_{\rm d}$   & $M_{\rm NS}$ & $P_{\rm orb}$& $\dot{P}_{\rm orb}$   & $\dot{M}_{\rm tr}$ & $\dot{M}_{\rm wind}$ & $R_{\rm d}$ & $T_{\rm eff}$ & $X_{\rm sur}$ & $\bar{\rho}_{\rm d}$\\
(d)&  (Gyr)& $(\rm M_{\odot})$ & ($\rm M_{\odot}$) & (0.01d) &($10^{-12}\rm s\,s^{-1}$)    & ($\rm M_{\odot}\,yr^{-1}$) & ($\rm M_{\odot}\,yr^{-1}$) & ($\rm R_{\odot}$)& (K) & &($\rm g\,cm^{-3}$)\\
 \hline\noalign{\smallskip}
2.3&9.61&0.015& 1.72 & 8.39& 2.19& 0                 & $0.3\times10^{-9}$ &0.093& 1877 &0.12&25.1\\
2.0&7.23&0.034& 1.87 & 8.39& 34.11&$2.5\times10^{-10}$& $6.0\times10^{-9}$&0.123& 2340 &0.48&25.6\\
1.5&3.63&0.041& 1.87 & 8.39& 4.32&$2.7\times10^{-11}$& $1.1\times10^{-9}$&0.131& 2326 &0.63 &25.9\\
1.2&2.18&0.043& 1.87 & 8.39& 4.19&$2.7\times10^{-11}$& $1.1\times10^{-9}$&0.132& 2324 &0.67 &26.1\\
1.0&1.54&0.044& 1.87 & 8.39& 4.34&$2.8\times10^{-11}$& $1.1\times10^{-9}$&0.133& 2319 &0.68&26.0 \\
0.8&1.12&0.044& 1.87 & 8.39& 4.35&$2.9\times10^{-11}$& $1.1\times10^{-9}$&0.134& 2317 &0.69&25.9 \\
\noalign{\smallskip}\hline
\end{tabular}
\end{minipage}
\end{table*}

In Table 2, we summarize the calculated relevant parameters for various initial orbital periods when the orbital period of LMXBs can fit the observed value of SAX J1808.4-3658. It is clear that, six cases that the initial orbital periods are below the bifurcation period can reproduce the orbital period of SAX J1808.4-3658. One can see in Table 2 that the larger initial orbital period (besides 2.3 d) would produce a higher orbital period derivative, higher mass transfer rate, higher wind loss rate, and smaller donor star mass. In particular, four cases when $P_{\rm orb,,i}=0.8$, 1.0 , 1.2, and 1.5 d can produce the observed orbital period derivative. Meanwhile, our simulations imply that the donor star of AMSP is a brown dwarf with 0.13 $\rm R_{\odot}$ and 0.044 $\rm M_{\odot}$. This mass is consistent with the minimum companion mass ($0.042\rm M_{\odot}$ at an inclination angle $i=90^{\circ}$) when $M_{\rm NS}=1.4~\rm M_{\odot}$ \citep{bild01}. However, because of the lack of dips or eclipses in the light curve, the inclination angle of this source is most likely below $60-70^{\circ}$ (see also the discussion in Section 4). Based on the accretion disk instability theory, the accretion rate of the pulsar is $\dot{M}_{\rm NS}=-\dot{M}_{\rm tr}/2d\approx 1.4\times 10^{-9}~\rm M_{\odot}\,yr^{-1}$ during outbursts. The peak X-ray luminosity of the pulsar is $L_{\rm X}=0.1\dot{M}_{\rm NS}c^{2}\approx8.0 \times10^{36}~\rm erg\,s^{-1}$, in approximately agreement with the observation. Furthermore, LMXBs in six cases would evolve into detached system, in which the donor star masses locate the definition of black widows ($0.02-0.05~\rm M_{\odot}$, Fruchter, Stinebring \& Taylor 1988; Stappers
et al. 1996).  Meanwhile, the mean density of the donor stars can exceed $23~\rm g\,cm^{3}$, which has found in the diamond planet PSR J1719-1438 \citep{bail11,benv12}.
\begin{figure}
\centering
\includegraphics[width=\linewidth,trim={30 0 90 30},clip]{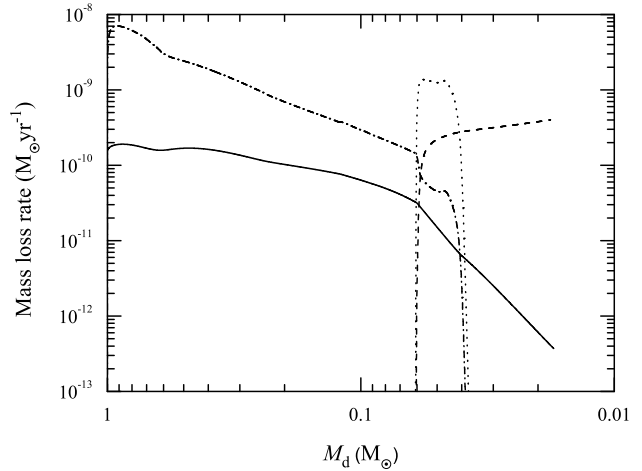}
\caption{Evolution of different mass loss rates from the donor star for an LMXB with a donor star mass of $1.0~\rm M_{\odot}$ and an initial orbital period of 1.0 day. The solid, dashed, and dotted curves represent the mass loss rate induced by angular momentum losses caused by gravitational radiation, evaporation process, and irradiation process, respectively. The dashed-dotted curve represent the mass transfer rate.} \label{fig:orbmass}
\end{figure}

In Figure 5, we present the evolution of $\dot{P}_{\rm orb}$ during the period-increasing phase in the $\dot{P}_{\rm orb}-P_{\rm orb}$ diagram. To distinguish the different curves, the evolutionary tracks have been slightly smoothed. Our scenario predicts that the orbital period derivative of SAX J1808.4-3658 has a tendency to decrease. However, the recent observation during the 2011 outburst of SAX J1808.4-3658 proposed that its orbital period derivative was rapidly increasing \citep{patr12}. If our simulation is near the truth, it is possible that their measured orbital evolution is just a short-term event.

\section{Discussion and Summary}

Employed the accretion disk instability model, the evaporation and irradiation wind model, in this work we have investigated the formation process of the first discovered AMSP SAX J1808.4-3658. The numerical calculations show that our evolutionary scenario can account for the observed orbital period,  the orbital period derivative, the minimum donor star mass, and the peak X-ray luminosity during outbursts.  Especially, an evaporation efficiency of 0.3 and a relatively higher irradiation efficiency of 0.1 can yield a high wind loss rate of $\sim10^{-9}~\rm M_{\odot}\,yr^{-1}$ from the donor star. If such a wind loss ejected from the inner Lagrangian point, the resulting orbital period derivative is $\dot{P}_{\rm orb}=(4.19-4.35)\times 10^{-12}~\rm s\,s^{-1}$, in approximately agreement with the observed value. Our simulations also show that SAX J1808.4-3658 would evolve into a detached binary, which resemble black widows or the diamond planet PSR J1719-1438. A detailed investigation for
black widows and the diamond planet is beyond the scope of this work, and will be achieved in a future work.

\begin{figure}
\centering
\includegraphics[width=\linewidth,trim={30 0 90 30},clip]{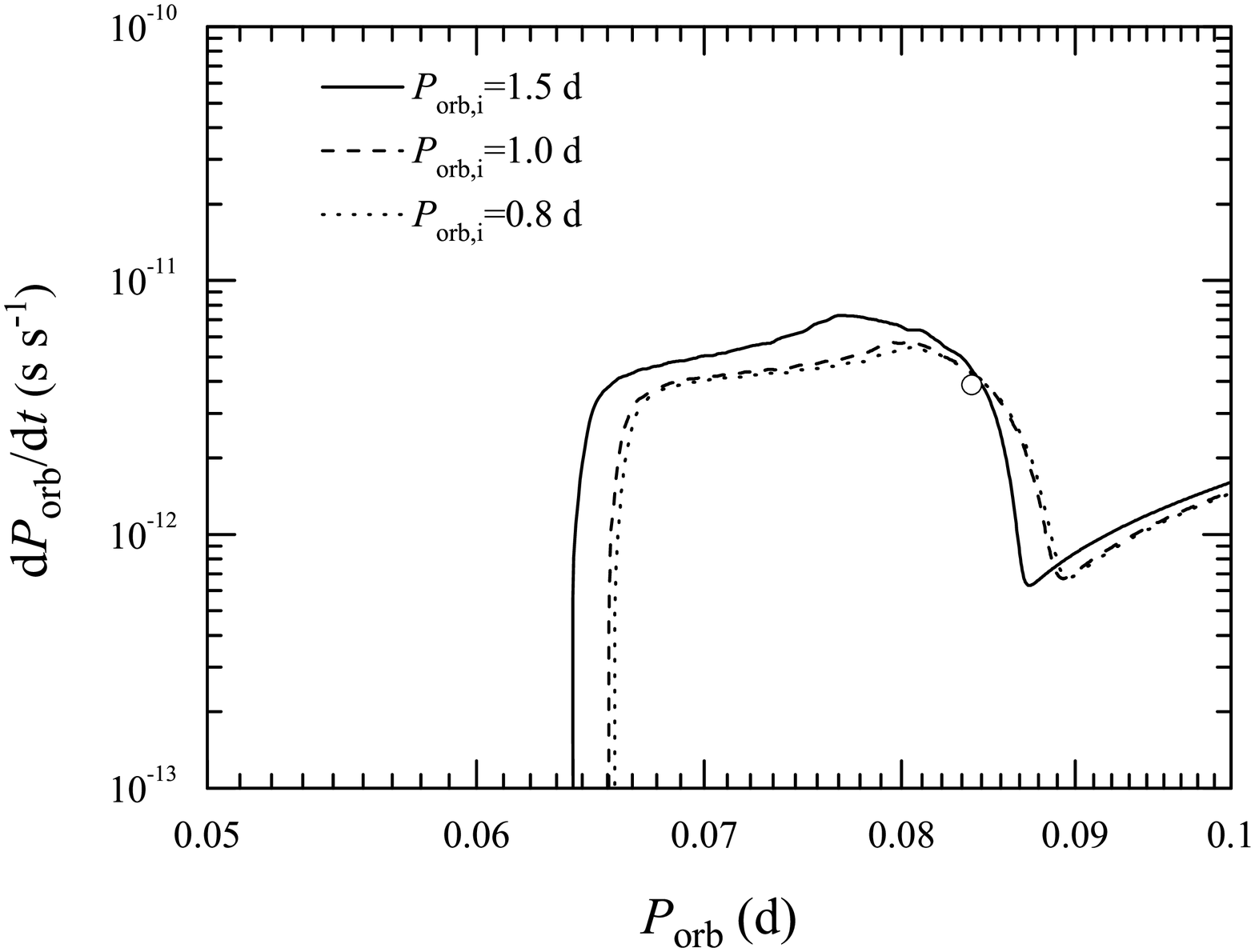}
\caption{Evolutionary tracks of LMXBs with a donor star mass of $1.0~\rm M_{\odot}$ and different initial orbital periods in the $\dot{P}_{\rm orb}-P_{\rm orb}$ diagram. The open circle represents the present parameters of SAX J1808.4-3658. } \label{fig:orbmass}
\end{figure}
Comparing with a relatively massive donor star (0.05-0.1 $\rm M_{\odot}$) proposed by \cite{bild01} and \cite{delo08}, our numerical simulation predicts a relatively small donor star mass of $\sim 0.044~ \rm M_{\odot}$. Because this source has not been reported the detections of dips and eclipses, its orbital inclination angle should be less than $60-70^{\circ}$. If our simulation is near the actual formation process of SAX J1808.4-3658, the pulsar should have a relatively low mass $M_{\rm NS}=1.32~\rm M_{\odot}$ ($i=70^{\circ}$) according to its mass function $f_{\rm x}=(M_{\rm d}{\rm sin}i)^{3}/(M_{\rm NS}+M_{\rm d})^{2}=3.8\times 10^{-5}~\rm M_{\odot}$ \citep{bild01}. \cite{di08} ideally explained the high orbital period derivative of SAX J1808.4-3658, while they only employed an analytical method to investigate its current evolution. Our numerical calculations can not produce such a high mass transfer rate ($\sim10^{-9}~\rm M_{\odot}\,yr^{-1}$), so a strong wind loss by evaporation and irradiation process is invoked to enhance the mass loss rate (it includes the mass transfer rate and the wind loss rate) of the donor star. Moreover, both approaches have a slight difference in the donor star structure: their mass-radius index is -1/3, but our index fitting by the donor star model in the range of 0.03 $-$ 0.06 $\rm M_{\odot}$ is about -0.2 (in MESA code, the stellar radius is calculated by a one-dimensional stellar evolution module according to its mass). Therefore, the donor star mass we obtained is obviously lower than their result (see also figure 2 of Di Salvo et al. 2008).

The standard theory thought that magnetic braking should stop when the donor star evolve to be fully convective \citep{skum72,verb81,rapp83}. However, some relevant works suggested that magnetic braking can not completely cease for stars without radiative core \citep{king02,andr03,pret08}. Recently, the orbital period of cataclysmic variable V2051 Oph with a 0.15 $\rm M_{\odot}$ secondary was found to be decreasing at a rate $\dot{P}_{\rm orb}=-5.93\times 10^{-10}~\rm day\,yr^{-1}$, which is about six times that induced by gravitational radiation \citep{qian15}. As a conclusion, \cite{qian15} proposed that this source supports the argument that full convective stars are undergoing magnetic braking. Our work also shows that magnetic braking without cut-off may play an important role in forming SAX J1808.4-3658. We also note that the irradiation efficiency ($f_{\rm ir}=0.1$) in our model may be slightly high. However, \cite{iben95} argued that the irradiation-driven wind loss rate is typically one order of magnitude higher than the mass transfer rate (see also their equation 25). Our mean wind loss rate also presentd a similar results (see also Figure 4). Therefore, our scenario combining evaporation wind and irradiation wind provides an evolutionary channel towards the AMSP SAX J1808.4-3658.

It is worth noting that, the actual mechanism resulting in the orbital expansion in SAX J1808.4-3658 should be more complicated than that given by our work. \cite{patr12} found that the orbit expansion of this source was accelerating a rate $\ddot{P}_{\rm orb}=1.6\times 10^{-20}~\rm s\,s^{-1}$. The orbital period derivative of SAX J1808.4-3658 remained stable in the past ten years \citep{burd09}, while the current observed timescale for the second orbital period derivative given by \cite{patr12} may be not enough yet to support a secular evolution. There exist a possibility that the observed orbital period derivative change is a short-term phenomenon, in which the angular momentum exchange by coupling between the donor star spin and the orbital motion. For example, a variation in orbital period may stem from a change of the mass quadrupole \citep{appl87,rich94,appl94}, which may originate from a deformation of the donor star by a strong magnetic activity \citep{appl92}.

\section*{Acknowledgments}
We are grateful to the anonymous referee for constructive comments. This work was partly supported by the National Science Foundation of China (under grant number 11573016), the Program for Innovative Research Team (in Science and Technology)
in University of Henan Province, and the China Scholarship Council.

\bsp

\label{lastpage}

\end{document}